\documentclass[12pt]{article}
\usepackage{graphicx,epsfig}
\usepackage{amsmath,amssymb}
\usepackage{latexsym}
\usepackage{amstext}
\usepackage{array}
\usepackage{multirow}
\usepackage{tikz}
\usepackage{ulem}
\usetikzlibrary{arrows}
\usepackage{amssymb}

\usepackage[ruled,vlined]{algorithm2e}

\newcommand{\mb}[1]{ \mbox{\boldmath$#1$} }
\newcommand{\ds}{\displaystyle}
\newcommand{\beq}{\begin{eqnarray}}
\newcommand{\eeq}{\end{eqnarray}}
\newcommand{\beqq}{\begin{eqnarray*}}
\newcommand{\eeqq}{\end{eqnarray*}}

\newcommand{\eps}{\varepsilon}
\newcommand{\x}{\mbox{\boldmath$x$}}
\newcommand{\X}{\mbox{\boldmath$X$}}
\newcommand{\Y}{\mbox{\boldmath$Y$}}
\newcommand{\f}{\mbox{\boldmath$f$}}

\newcommand{\Z}{\mbox{\boldmath$Z$}}

\newcommand{\m}{\mbox{\boldmath$\mu$}}
\newcommand{\llll}{\mbox{\boldmath$\lambda$}}
\newcommand{\s}{\mbox{\boldmath$s$}}
\newcommand{\mm}{\mbox{\boldmath$m$}}

\begin{document}
\title{Single particle algorithms to reveal cellular nanodomain organization}
\author{P. Parutto$^{1}$, J. Heck$^{2}$, M. Heine$^2$ and D. Holcman$^{3 ,4}$ \footnote{$^{1}$ UK Dementia Research Institute at the University of Cambridge and Department of Clinical Neurosciences, University of Cambridge, Cambridge CB2 0AH, UK. $^2$ Research Group Functional Neurobiology at the Institute of Developmental Biology and Neurobiology, Johannes Gutenberg University Mainz, Mainz, Germany   $^3$ DAMPT, University Of Cambridge, DAMPT and Churchill College CB30DS, United Kingdom. $^{4}$ Group of Data Modeling and Computational Biology, IBENS, Ecole Normale Sup\'erieure,75005 Paris, France.  Lead contact: david.holcman@ens.fr}}
\date{\today}
\maketitle
\begin{abstract}
Formation, maintenance and physiology of high-density protein-enriched organized nanodomains, first observed in electron microscopy images, remains challenging to investigate due to their small sizes. However, these regions regulate molecular trafficking, assembly and sorting required for higher cell functions, such as communication or plastic changes. Over the past ten years, super-resolution single-particle trajectories (SPTs) have been used to sample these sub-cellular environments at a nanometer resolution for both membrane and soluble proteins. We present here data analysis developments and algorithms that convert high-throughput molecular trajectories into maps of molecular density, diffusion and local drift organization. These approaches transform intrinsic trajectory properties into statistics of the underlying cellular organization. The automatic identification of large numbers of high-density regions allows quantifying their boundary location and organization, their stability over time and their ability to transiently retain molecules. To conclude recent automated algorithms can now be used to extract biophysical parameters of sub-cellular nanodomains over a large amount of trajectories.
\end{abstract}
\maketitle
\section{Introduction}
High-density nanodomains in neurons and in cells in general have already been characterized from the early ultrastructure-electron microscopy images of Palade, de Robertis, and Bennett in 1954, suggesting more than 70 years ago \cite{bennett1956concepts} the heterogenous distribution of proteins especially at neuronal synapses such as post-synaptic density (PSD). This observation anchored the field of neurotransmission at the molecular organization level, where local structures shape function.\\
Dense nano-regions are ubiquitous on membranes such as the pre-synaptic active zones that are enriched in voltage-gated Calcium channels (VGCCs) and docking proteins and serve as priming sites for readily releasable synaptic vesicles \cite{heck2019transient}. Organelles such as the spine apparatus (SA) \cite{segal2010spine,korkotian2014synaptopodin}, a specialized compartment of the endoplasmic reticulum (ER), found in a subpopulation of dendritic spines in neurons of the central nervous system is also organized in nanodomains with differential accumulation of Ryanodyne receptors (RyR) at the base versus SERCA pumps located inside the spine head, thus regulating calcium in and out fluxes \cite{basnayake2021,basnayake2019fastplos}. Made of molecular aggregates, these nano-regions are also observed in soluble environments, for example inside the nucleus with high density-regions enriched in nucleosomes, or in the cytosol where reserve and recycling pools of synaptic vesicles are transiently accumulated in axons and synapses \cite{gormal2020modular,gormal2022nanoscale}.\\
Characterizing and identifying high-density nandomains remained challenging as they are formed by dense protein assemblies, measuring only a few hundred of nanometers. Recent findings and reviews \cite{Zhang2016,wu2020liquid,pak2016sequence} have proposed that these regions could result from spontaneously organized condensed phase where higher protein concentration accumulate, with the classical image of oil droplets in water. This representation is not as simple, as it consists in multimeric assemblies of various types of interacting proteins. The concept of phase separation is indeed a physical model to describe membraneless compartments also known as molecular condensates. We refer to recent reviews on phase separation in synaptic biology for these phase separation concepts \cite{wu2020liquid}.\\
However, in the past decade, single particle trajectory (SPT) approaches revealed that these condensates are permissive to proteins, and thus are not necessarily creating fully isolated domains. In addition these regions are not necessarily stable over time. We present here how high-density nanodomains have been characterized by SPTs, their temporal stability, and how they can retain molecules over time. We further develop ready-to-use automated algorithms, developed to detect high-density regions from SPTs analysis and to collect their statistics.
\section{Sub-cellular space exploration by SPTs}
Single-particle stochastic trajectories are now routinely collected to explore sub-cellular environments, such as the spatio-temporal organization of neuronal \cite{heck2019transient}, immune cells \cite{ponjavic2018single,dustin2012receptor,santos2018capturing} or organelles \cite{Parutto2018}. These trajectories for many molecules are generated by methods such as sptPALM or UPAINT, allowing to retrieve the organization of the local environment (cytoplasm, membrane) that is explored by individual molecules. They can also serve to explore protein-receptor interactions occurring between G protein-coupled receptors and G-proteins \cite{calebiro2021g,sungkaworn2017single} or interactions between endogenous adrenoreceptors in neurosecretory cells \cite{gormal2020modular}.\\
Traditionally, SPT analysis relies on Mean-Squared Displacement (MSD), a statistical estimator allowing to roughly characterize the motion of trajectories as sub-, super- or diffusive.  In the case of a diffusive motion, it retrieves the diffusion coefficient \cite{saxton1997}. The MSD however is based on averages that heavily limit their use for resolving spatio-temporal variations in the particles motion. In the past years, novel estimators have been developed that rely on local averaging \cite{Hoze2017}, thus allowing to keep the spatio-temporal heterogeneities of the cells \cite{parutto2022high}. These approaches allowed to shift from the global analysis of individual trajectories to the local analysis of specific regions combining many trajectories, thus retrieving the properties of the underlying explored regions \cite{holcman2015analysis}.\\
For example, the association (on rate) $k_{on}$ values for receptor G protein interactions was recently measured by single-molecule microscopy, and was found to be 10 times higher for the a 2A-adrenergic receptor with Gai than for the beta 2-adrenergic receptor with Gas. Receptors and G proteins stayed together for about 1-2s, leading to a $k_{off}$ of 0.5-1s\cite{calebiro2021g,sungkaworn2017single}. The on rate $k_{on}$ has not been measured so far from SPts on membrane receptors inside nanodomain, as it requires to delimit the region (which could be possible with the methods reviewed here), but requires enough trajectories that would pass inside. In general, the forward rate $k_{on}$ can be evaluated as the mean time to reach a region starting from another one \cite{HozeBJ2014}.
\begin{figure}[http!]
\begin{center}
 \includegraphics[scale=0.8]{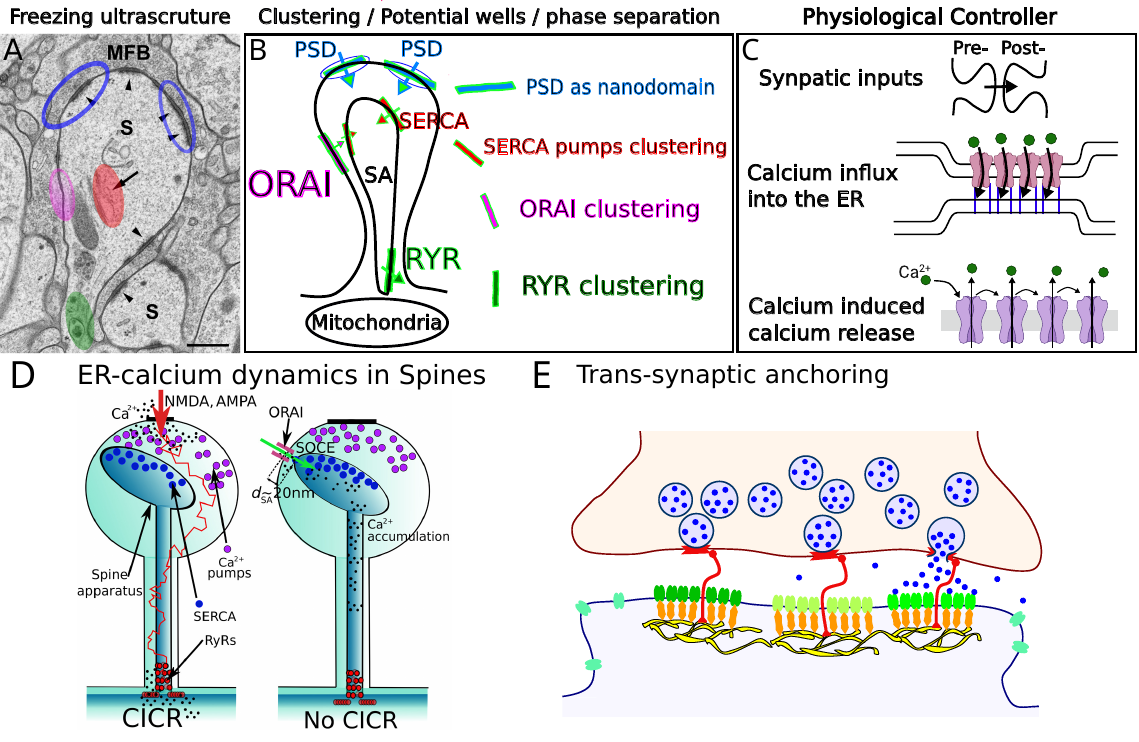}
\caption{ {\bf Nanodomains organization and function.} {\bf A.} Freezing dendritic spine ultrastructure with various colored coded nanoregions. {\bf B.} Example of a dendritic spine's (with a spine apparatus) local molecular organization with SERCA pumps, Ryanodyne receptors (RyR), ORAI clusters and the post-synaptic density (PSD). {\bf C.} Nanodomain functions: synaptic function allows PSD to regulates synaptic influx, ORA-STIM1 regulates calcium ER replenishment. SERCA pumps are clustering, which is responsible for pumping calcium in the spine apparatus. RyR cluster modulates calcium-induced-calcium-release to control mitochondria ATP production. {\bf D.} Left: fast calcium transmission from the spine head to the base, leading to calcium spine apparatus depletion, mediated by the first arriving calcium (extreme statistics) \cite{basnayake2018fastest}. Right: slow calcium influx resulting in SA replenishment, adapted from \cite{basnayake2021}.  {\bf E.} Nano-column organization of the synaptic cleft to increase the communication speed and efficacy, adapted from \cite{freche2011}.}
\label{figureintro}
\end{center}
\end{figure}
\section{Properties of nanodomains}
High-density nanodomains have been revealed by electron microscopy (Fig.~\ref{figureintro}A) and by overlapping single stochastic trajectories and appear in a large diversity of environments. These domains play a fundamental role in transiently stabilizing trafficking elements such as proteins, vesicles, cargoes, etc at key locations (Fig~\ref{figureintro}B) allowing fine-tuning of the spatio-temporal cell physiological responses (Fig.~\ref{figureintro}C).
\subsection{Post-synaptic density nanodomains}
The stabilization of post-synaptic receptors at post-synaptic densities (PSD) is a fundamental characteristic to optimize synaptic transmission and to reduce the need of too many dendritic spines \cite{freche2011,heine2020asymmetry}. PSDs have been proposed to result from the spontaneous concentration of hundreds of molecular assemblies, scaffold proteins such as PSD-95, GKAP, Shank, Homer and CamKII \cite{nicoll2022} suggesting that they could form a phase separation with high concentrations of molecular components without needing physical barriers to separate the condensed phase from the rest of the membrane (bulk phase). However, this possibility has to be reconciliated with the constant exchange of AMPA or NMDA receptors or Glycine or GABA receptors, which would usually be prevented to enter these condensate in a phase separation \cite{chen2020phase}.\\
PSD identity seems to be maintained by the number of CaMKII proteins, a guardian keeper of synaptic efficacy, probably resulting in the local organization of the PSD. As CamKII needs to be replaced, the PSD should allow new CAMKII to come close to mature ones to activate them by phosphorylation \cite{lisman2012mechanisms,nicoll2022}. This cycle should keep the number of activated CaMKII constant. But this stabilization again requires access to the PSD.
\subsection{Nanodomains of receptor cluster to robustly and efficiently transmit fast signal}
Molecular clustering in cell can play multiple roles: 1- a high number and concentration guarantee a robust signal transmission; 2- Create an avalanche when the messenger is also released; 3- The cluster can be position at a key position to amplify a signal. We shall describe here some key cluster organization at dendritic spines (Fig. \ref{figureintro}A): Ryanodine receptors (RyRs) are present at the surface of the Endoplasmic Reticulum (Fig. \ref{figureintro}B), where they can organize in clusters \cite{jayasinghe2018true}, especially on the SA at the base of dendrites, resulting in calcium-induced calcium release (CICR), a process involved in large calcium releases activating mitochodria \cite{basnayake2019fastplos}. The local structural organization is designed to underlie a function: while receptor clusters at the PSD guarantee a robust synaptic transmission, ORAI proteins \cite{wu2014single} form clusters at the head of dendritic spines close to the Spine Apparatus where Sarco/Endoplasmic Reticulum Calcium ATPase (SERCA) pumps are located, possibility to guarantee an efficient refilling of the ER calcium (Fig.~\ref{figureintro}C-D) \cite{basnayake2021}. SPTs of STIM and ORAI proteins previously revealed \cite{wu2014single} that they transiently interact to form a confined nanodomains (Fig. \ref{figureintro}C). Similarly, IP3 receptors can rapidly and reversibly form small clusters of $0.5 \mu m$ containing tens of receptors, by a yet unknown mechanism, that could involve diffusion \cite{rahman2012dynamic, thillaiappan2017}. The role of IP3 receptor clustering is probably to increase CICR probability.
To conclude, protein clustering especially at synapses generates a robust signaling, as shown here with the case of calcium fluxes. The structures, modeling and re-organization of these nanodomains remain unknown.
\subsection{Nanodomains organization at the NMJ synapses}
Nanoregions can also appear organized at micrometer level. This is the case at the neuro-muscular junction (NMJ), where PSDs are mostly located in these regions, This is in contrast with Calcium voltage channels at pre-synaptic terminals of hippocampal neurons \cite{heck2019transient}. Nanoregions are organized on elliptical domains surrounding the NMJ as recently demonstrated \cite{ghelani2023interactive}.
\subsection{High density nanodomains and co-aligned nanodomains}
While constant reorganization of signaling molecules within pre- and post-synaptic compartments sustain synaptic transmission and plasticity, a form of stability is maintained by alignment of pre- and post-synaptic terminals. Indeed, voltage-gated calcium channels (VGCCs) intermingle with docked vesicles forming dense domains, the postsynaptic side is well organized to accommodate ligand-gated ion channels such as AMPA receptors (AMPARs), NMDAs at excitatory synapse \cite{biederer2017} or glycine and GABA receptor at inhibitory ones. Interestingly, pre- and post-nanodomains can be further co-organized in nanocolumn \cite{freche2011,heine2020asymmetry} that originates from the alignment of pre- and postsynaptic scaffold proteins, so that VGCCs and post-synaptic receptors are distributed in an optimal configuration that guarantee that vesicular release sites in the presynaptic terminal can activate the maximum number of receptors on the postsynaptic site \cite{heine2020asymmetry} (Fig.~\ref{figureintro}E).
\subsection{Nanodomains: exclusion phase separation or attractor potential wells}
The nature of nanodomains remains difficult to apprehend: they should capture receptors for a transient time, but remains permissive to guarantee a balance between incoming and leaving receptors. These nanodomains result
from interaction-mediated molecular networks with possible co-assemblies of positively charged partners to form dense liquid droplets \cite{wu2020liquid,pak2016sequence}. However SPTs reveals a different signature associated with long-range interaction of hundreds of nanometers, that could reflect a hidden organization \cite{hoze2012heterogeneity,Hoze2017}. \\
These nanodomains have been characterized as trapping structure such as potential well. Similar to a local chemical, mechanical or electrical interactions \cite{Schuss2,Schuss4}, a potential well generates long range interactions inside its region up to its boundary. The strength of the interaction is measured by the depth of the well, in the energy units of kT. An energy of 1-2 kT is considered to be inside the thermal noise, 3-4 kT are small energies and $>5$ kT can be consider high energies. Proteins considered as stochastic particles can be trapped in a potential well. Escape from a potential is possible by thermal noise to overcome the deterministic force which pushes back the stochastic particle toward the well center \cite{Schuss2,Schuss4}. For classical Brownian motion, the mean residence time which characterizes the strength of the well depends on the energy barrier and diffusion coefficient \cite{Holcman2015}. It could be possible that the well boundary is not necessary fully permeable to trajectories (partial absorption) \cite{Holcman2015}, a situation that increases the residence time, as trajectories could only escape through small openings. Local interactions would also retain trajectory longer \cite{taflia2011estimating}. To conclude, a class of nanodomains can be described as resulting from a long-range  interaction, modeled by potential wells. In the remaining part of the review, we will summarize modeling, algorithms and software pipeline (fig. \ref{figurePipeline}) to reconstruct these potential wells from SPTs.
\section{From SPTs to nanodomain reconstruction using biophysical modeling}
\begin{figure}[http!]
\begin{center}
\includegraphics[scale=0.6]{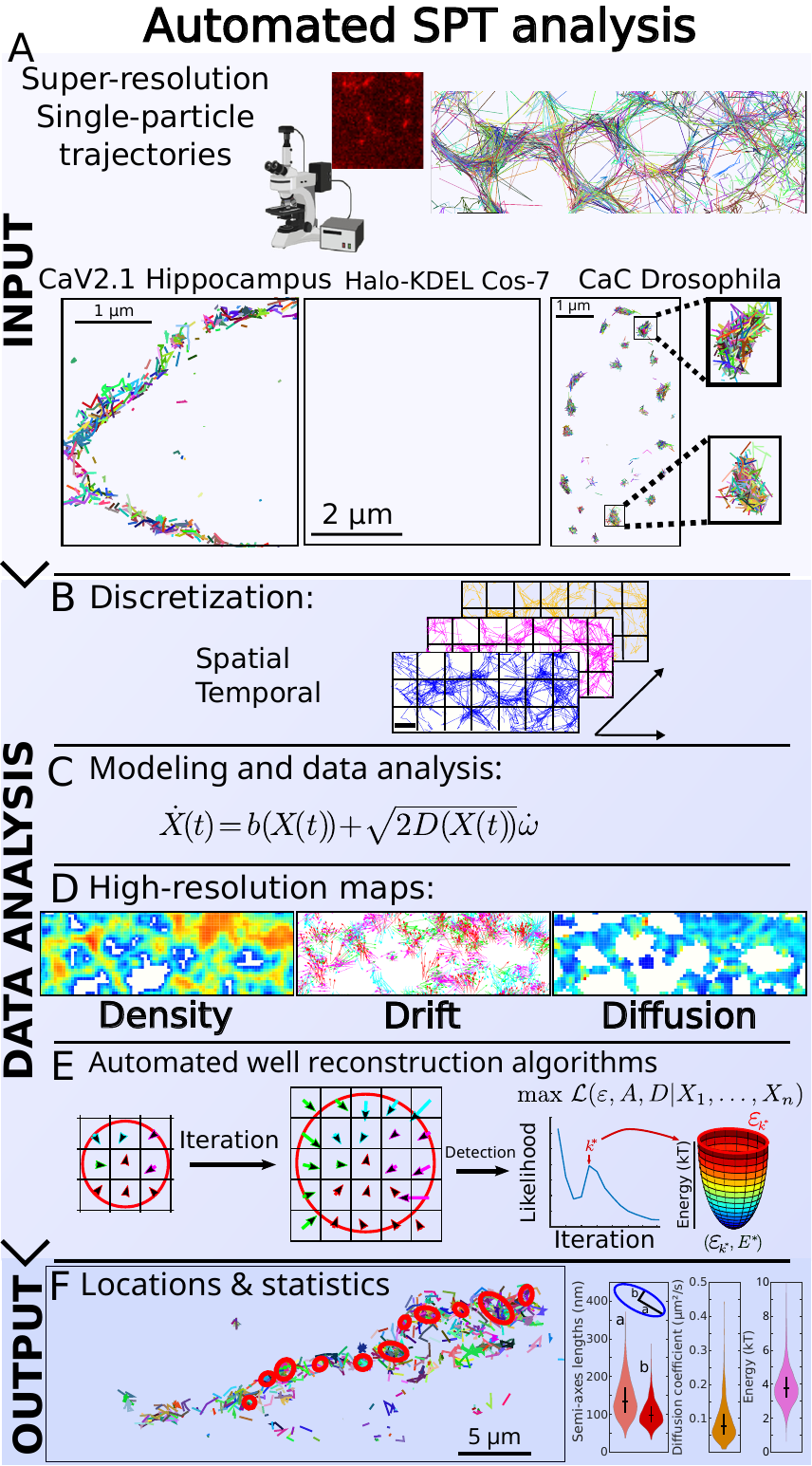}
\caption{ {\bf High throughput single particle trajectory pipeline.} (A) SPT input recorded from super-resolution microscopy tracked through time using a tracking algorithm to generate trajectories from successive points. Examples of trajectories from CaV2.1 calcium channel in hippocampal neurons, ER luminal probe in COS-7 cells, and CaC channels at Drosophila NMJ. (B) Discretization of the trajectories using temporal time-windows and spatial grids. (C) Overdamped stochastic Langevin equation used to interpret the trajectories. (D) Density, diffusion and drift maps for interpreting the local dynamics. (E) Algorithms to extract potential wells and statistical information about nanometer high-density region. (E) Outputs of the method: localization of the detected wells and their associated population statistics.}
\label{figurePipeline}
\end{center}
\end{figure}

SPTs appear as a high precision tool to explore nanodomain at the molecular level. Tagged receptors can even interact thus revealing local interactions. Over the past tens years, various data-driven methods have been developed to reconstruct nanodomains as a reverse engineering problem. We review here these approaches based on stochastic modeling at the base of high throughput data, pipeline and automated algorithms to reconstruct the biophysical properties of nanodomains based on single particle trajectories (Fig. \ref{figurePipeline}). Automated SPTs analysis starts by collecting data from super-reolution microscopy to generate trajectories (Fig. \ref{figurePipeline}A). Then, a Biophysical model defines the parameters to be extracted (Fig. \ref{figurePipeline}B-C) with a given spatio-temporal resolution. These computations allow to define two-dimensional maps for the density, diffusion, drift as well as the properties of high density regions that define nandomains (Fig. \ref{figurePipeline}D-E). a potential well model is used to interpret high-density regions, a signature of which is a locally converging drift vector field (Fig.~\ref{figurePipeline}E).
\subsection{Molecular motion described by Diffusion models}
The molecular motion in cells, for both soluble and membranes proteins, is driven by a mix of random and deterministic forces. While random forces results from the thermal fluctuations, deterministic ones can results from local electrical forces, mechanical, membrane curvature or flows. The position $\X(t)$ of a particle at time $t$ is well described by Langevin's equation \cite{Einstein, Brush,Langevin}: it describe the motion of a stochastic particle driven by a random force $\eta$ (accounting for thermal fluctuations) and an external field of force $F(\X,t)$ (e.g., electrostatic, mechanical, etc). When the force field is locally the gradient of a potential, we have $F(\X,t)=- \nabla U(\X)$. In the large friction limit ($\gamma \gg 1)$, which is the case in liquids and membranes, the Langevin equation becomes the Smoluchowski equation \cite{Schuss2,Belopolskaya} given by
\beq
\gamma \frac{d\X}{dt}=-\nabla U(\X)+\sqrt{2\eps\gamma}\ \dot{\eta}.
\label{LSE12}
\eeq
where the energy is $\eps=k_B T$. For timescales much longer than elementary molecular collision events, eq. \ref{LSE12} \cite{Saffman1975,Langevin} is replaced by the coarser effective stochastic equation \cite{HozePNAS,HozeBJ2014}
\begin{align}\label{stochlocal01}
\dot{\X}=\mb{a}(\X) dt +\sqrt{2} B(\X) \dot{W},
\end{align}
where  $\mb{a}(\X)$ is the overall drift field and $B(\X)$ is a smooth diffusion matrix related to  the effective diffusion tensor by  $D(\X)=\frac{1}{2} B(\X) B^T(\X)$ ($.^T$ denotes the transposition) \cite{Schuss4,Schuss2}. Fast variations of $B(\X)$ at small spatial distances should be resolved directly to avoid artifacts of nonphysical discontinuities that could corrupt the stochastic description. The parameters of model (eq.~\ref{stochlocal01}), the drift field $\mb{a}(\X)$ and the effective diffusion tensor $D(\X)$, can be estimated from trajectories \cite{HozePNAS,Hoze2017,parutto2022high}.
\subsection{Normal versus anomalous diffusion model}
The previous Smoluchowski description (eq.~\ref{LSE12}) breaks down when correlated random forces drive the movement of the observed particle. This is the case for chromatin loci or receptors evolving in a network of correlated scaffolding molecules \cite{Metzler2001,Amitai2017}. This effect is reflected in single particle trajectories. To test for such correlations, the statistics of the spatial increments $\Delta \X= \X(t+\Delta t)- \X(t)$ are computed. \\
The identification of correlated motion starts by averaging $\Delta \X$ along a trajectory or on many displacements falling at a given location (small domain around a point), followed by fitting, for small time lag $\Delta t$, the curve  $(\Delta t)^{\tau}$ to the second statistical moment of the displacement, leading to
\beq
\langle |\X(t+\tau)-\X(t)|^2\rangle \approx A \tau^\alpha,
\eeq
where $A$ is a constant, the motion is classified as Brownian when $\alpha=1$. For $\alpha>1$, the underlying dynamics is associated with large excursions (super-diffusion), which could be due to a combination of a drift and Brownian motion \cite{AmitaiCellreport2017}. When  $\alpha<1$, the displacements are restricted, similar to a locus attached to a chromatin \cite{AmitaiCellreport2017,Amitai2017,Shukron2017PRE,shukron2019}.\\
Modeling the motion of correlated particle with an exponent $\alpha$ inside a field of force uses the fractional Brownian motion (fbm), transforming eq.\ref{stochlocal01} into
\beq\label{eqstochastic}
\dot{\X}=\mb{a}(\X)+\sqrt{2}\sigma(\X) \dot{\mb{B}}_{H}(\X)\hspace{0.5em}\mbox{for}\ \X\in\Omega,
\eeq
where the sub-diffusive process $B_H(t)$  \cite{Mandelbrot1968} following the properties
\begin{eqnarray}
&\langle B_H(t)\rangle = 0 \\
&\langle B_H(t) B_H(s) \rangle = \frac{1}{2} \left( t^{2H} + s^{2H} - |t-s|^{2H} \right),
\end{eqnarray}
$2H=\alpha$.
In this case, recovering the field of force from SPTs becomes more difficult.
\subsection{Localization errors}
A possible source of error on the positions of the observed particles comes from the detection device. This results in a source of noise that is independent from thermal fluctuations and should be accounted for \cite{berglund2010,HozePRE2015}. This effect is modeled as a Gaussian with variance $\sigma$. Since the stochastic noise associated with the underlying physical process is independent from the localization noise, the measured MSD contains now two terms: the one associated with the localization error and the other due to physical correlation, leading to
\beq
\langle |\X(t+\tau)-\X(t)|^2\rangle \approx A \tau^\alpha +\frac{\sigma^2}{2}\tau.
\eeq
When the localization error is modeled by an Ornstein-Ulhembeck process $\Z(t)$, the measured signal is the sum  $\Y=\X+\Z$, where $\X$ is the physical motion. The dynamics of $\Z$ is $\dot{\Z}=-\lambda \Z + \sigma \dot{\omega}$ , where $\lambda$ and $\sigma$ are two parameters. The correlation is given by
$\langle|\Z(t+\tau)-\Z(t)|^2\rangle =\sigma^2\left( \frac{1-\exp (-2 \lambda t)}{2\lambda}\right)$.  When the correlation time is longer than the localization error relation time, we obtain $\langle |\X(t+\tau)-\X(t)|^2\rangle \approx A \tau^\alpha +\frac{\sigma^2}{2\lambda}$. \\
\subsection{Nanodomain described as a potential well or phase separated condensate} \label{ss:potentialwell}
A nanodomain consists of a high-density region, concentrating a mix ensemble of molecules. The dual ability to retain molecules while allowing a certain fraction to enter or to exit could rely on the local molecular organization.
This organization could generate long-range interaction or/and extended forces.  The drift field $\mb{a}(\X)$ in eq.~\ref{stochlocal01} can account specifically for long-range forces that acts on diffusing particles \cite{holcman2013unraveling}. However, short-range forces that would result in local cluster due to electrostatic interactions would require a more detail model to account for exclusion volume or Lennard-Jones forces \cite{chandrasekhar1943stochastic}. Inside nanodomains, the diffusion coefficient $D(\X)$ can be considered to be locally constant and the coarse-grained drift field $\mb{a}(\X)$ is the gradient of a potential
\beq
\mb{a}(\X)=-\nabla U(\X),
\label{welldrift}
\eeq
where
\beq \label{eqA}
U(\X)=A \left( \left(\frac{x-x_0}{r_x}\right)^2 + \left(\frac{y-y_0}{r_y}\right)^2\right).
\eeq
Here $r_x, r_y$ represent the two semi-axis lengths and the parameter $A$ measures the force of the field. The force acting on a particle trapped in a potential well is
\beq\label{eq:fbmparab}
\mb{F}(\X) = -\frac{2A}{r^2}(\X - \m),
\eeq
as long as the particle is inside the well $|\X - \m|<r$. As we shall see below, these three parameters can be identified from SPT data. Following this previous description, the density of molecules inside the nandomain follows the Boltzmann distribution $e^{-U(\X)/D}$ \cite{HolcmanBJ2015}. Note that other forms for the potential energy function are also possible such as
\beq
U(\X)=
\left\{\begin{array}{lll}
&\ds{A\left[\left(\frac{x-x_0}{r}\right)^{2k} + \left(\frac{y-y_0}{r}\right)^{2k}-1\right] } & \mbox{ for }  (x-x_0)^2+(y-y_0)^2 <r^2
\\ &&\\
& 0& \mbox{otherwise.}
\end{array}\right.
 \label{eqpotentialwell}
\eeq
where the parameter $k$ measures the flatness of the field inside the nandomain. We discuss in the next sections different estimators and algorithms for estimating potential wells.
\section{From parameter estimation to statistical maps}
Statistical estimators based on local SPTs averaging have been used to extract deterministic forces and the spatial dependent diffusion coefficient from eq. \ref{eqstochastic} and \ref{stochlocal01} \cite{HozePNAS,Hoze2017}. The general theory for constructing these estimators have been reviewed in \cite{schuss1980,Schuss2009,HolcmanBJ2015,Hoze2017} see also Box 1). Beyond Bayesian inferences \cite{hamilton2020time,prakasa1999statistical}, empirical estimators allow recovering various maps such as density, drift and diffusion but also identifying potential well parameters such as the depth and the boundary. In particular, the known limiting factors to the accuracy of potential well estimation are:
\begin{enumerate}
  \item Absence of trajectories covering the nanodomain boundary. This region is poorly sampled due to the large gradient at the boundary.
  \item Trajectories can bleach before escaping, leading to sub-sampling and thus large deviations compared to the exact structure of the boundary \cite{meiser2023experiments}.
  \item A large acquisition time step $\Delta t$ leads to large fluctuations in all these parameters.
  \item A small acquisition time step $\Delta t$ allows a mild spatial resolution improvement, limited by the length $\sqrt{2D\Delta t}$ of a random walk.
\end{enumerate}
\subsection{Increasing spatial resolution: from a fixed grid to a sliding window analysis}
\begin{figure}[http!]
	\begin{center}
		\includegraphics[scale=0.7]{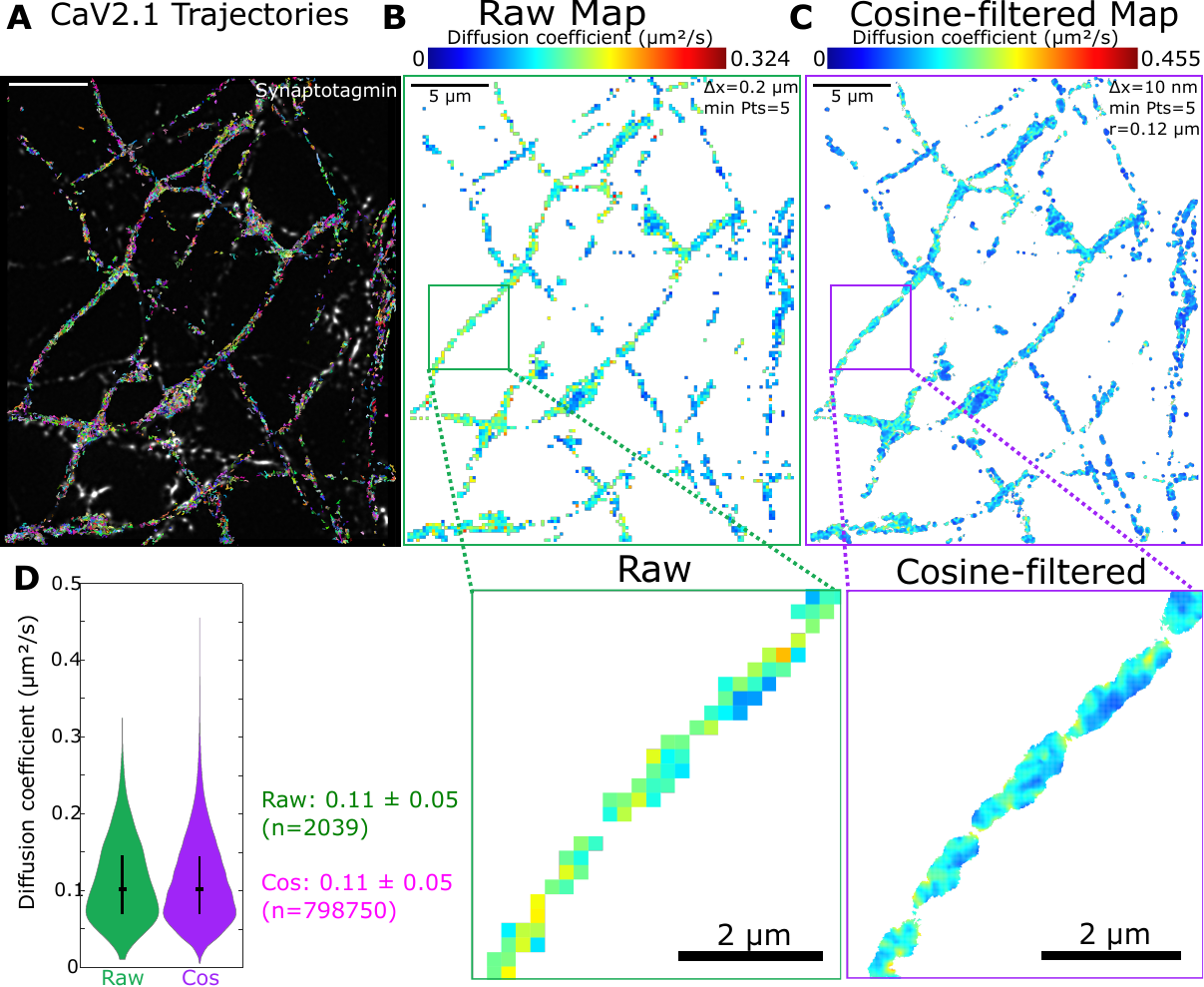}
		\caption{ {\bf Cosine-filtering to improved map accuracy.} {\bf (A).} Trajectories of Calcium Voltage channels (Cav2.1) at the surface of hippocampal neurons overlaid on top of the synaptotagmin image. {\bf (B).} Raw diffusion map obtained for a grid with $\Delta x = 0.2 \mu m$ computed on the trajectories presented in (A). {\bf (C).} Cosine-filtered diffusion map based on a grid with $\Delta x = 0.01 \mu m$ computed on the same trajectories as in B \cite{parutto2022high}. (D) Quantification of the recovered diffusion values for the two methods.}
		\label{figureCosFilt}
	\end{center}
\end{figure}

Computing parameters such as the density, diffusion or vector field map typically depend on a fixed grid that restricts the precision to the chosen size $\Delta x$. Indeed, the different quantities are computed for each bin of the grid and is limited by the number of trajectory points falling inside each bin \cite{hoze2012heterogeneity}. \\
To improve the accuracy of these maps, a general approach consists in decreasing the bin size $\Delta x$, but this reduces the number of points per bin, thus increasing the uncertainty. Another possibility is to compute any estimator on a sliding disk $D(P,\eps)$ centered at point $P$ and of radius $\eps$ \cite{parutto2022high}. The point $P$ can then moved in a much smaller grid of size $s=\eps/n$, where $n$ can be large $n=5,10,...$, so that the resolution is much smaller than $\Delta x$. In that case, the limiting resolution is $\eps$. \\
To avoid possible boundary effect, it is possible to multiply any estimator by a weighting function that depends on the distance to the center of the bin (Fig.~\ref{figureCosFilt}A-C) such as $\cos (r\frac{\pi}{2\eps})$ (Fig.~\ref{figureCosFilt}C). Other weighting methods can also be used, such as the Laplace filter averaging, but the cosine-filtering recovers more local details.
Interestingly the distribution of parameters such as the mean-square-displacement or diffusion does not vary much between a uniform grid and the sliding window approach.

In summary, this procedure allows to increase the resolution by a large factor but involves two scales: the radius $\eps$ for the moving disk, and the displacement $\Delta x_d$ on a defined grid. If there are not enough point falling in the disk, the corresponding bin can either be discarded or a larger disks such as $D(P,2\eps)$, $D(P,2^2\eps)$ can be used, but at the cost of coarse-graining the spatial resolution.
\subsection{Segmenting high density regions using Voronoi Tesselation}
\begin{figure}[http!]
\begin{center}
\includegraphics[scale=0.7]{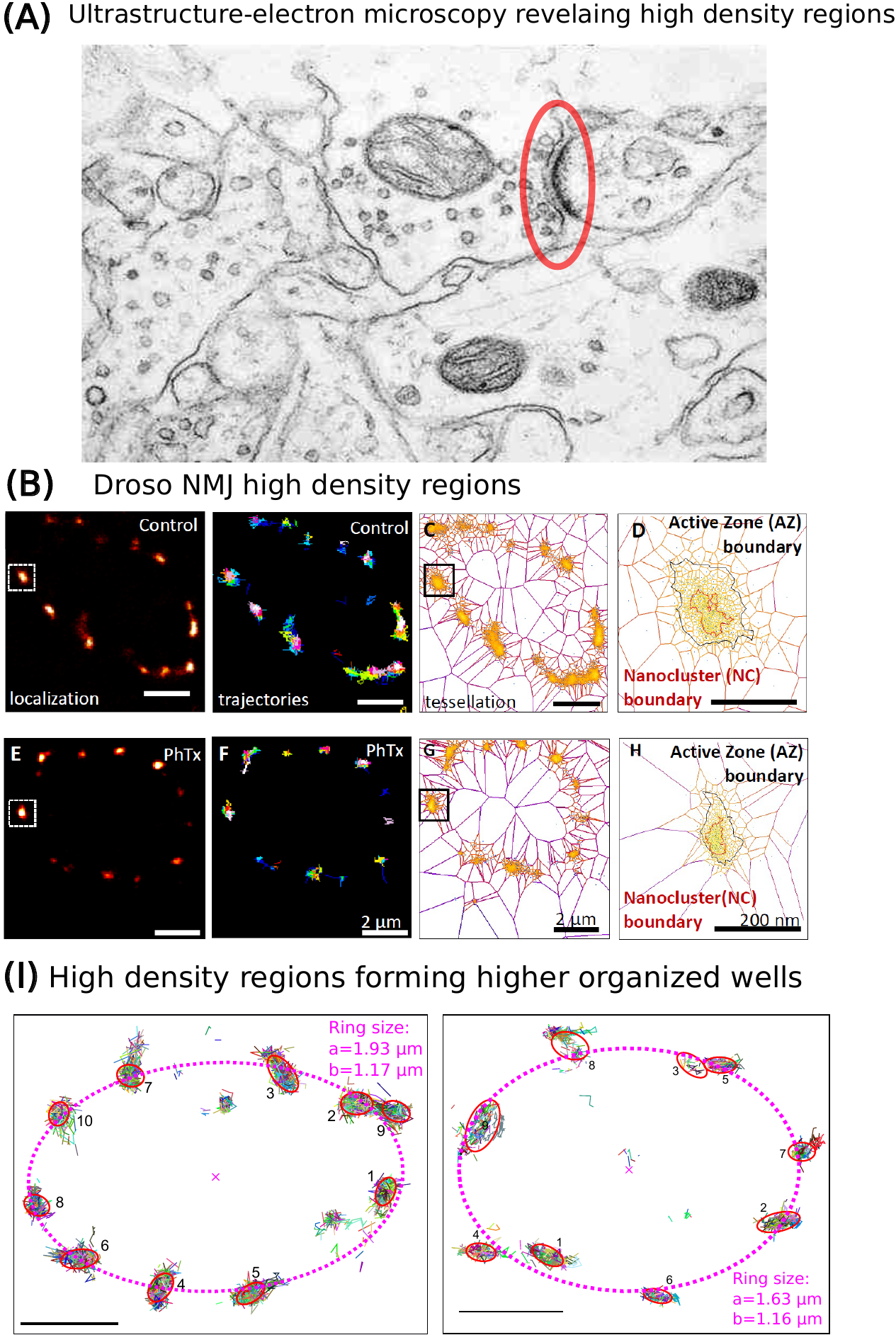}
\caption{ {\bf From ultrastructure to nandomain organization.} (A) Ultrastructure historical image (Blobel 1950) showing vesicle organization around a synapse.
{\bf B-H.} Live sptPALM imaging of CacmEOS4b at muscle 4 or 6/7 NMJs, performed in 1.5 mM Ca2+ 5 min after a 10-min incubation in HL3 with PhTx or in plain HL3 (Control). Images show representative PALM recordings (B and E), trajectory maps (B and F), and tessellation analysis representations of CacmEOS4B (C, D, G, and H). Scale bars, $2 \mu m$ (A to C and E to G) and 200 nm (D and H). {\bf I.} Two regions showing potential wells organised in rings detected in trajectories of Cac channels in third instar CacmEOS4B larvae localized at AZs. Scale Bar $1 \mu$ m. Adapted from \cite{ghelani2023interactive}.}
\label{figureNanodomains}
\end{center}
\end{figure}

To go beyond the ultra-structure images of neurons (Fig.~\ref{figureNanodomains}A), it is possible to identify protein clusters and their boundary according to the static density distribution of points obtained from single-molecule microscopy\cite{levet2015sr}, as illustrated for the calcium voltage channel for the neuromuscular junction (Fig.~\ref{figureNanodomains}B). Points that form cluster can be extracted from the Density-based Spatial Clustering Analysis (dbscan) algorithm \cite{parutto2022high}, which separate clustered points from isolated ones based on whether the number of neighbors in a disk of radius $\epsilon$ exceed a thresholds $T$. Another possibility is to use the K-Ripley function \cite{lagache2022} to determine whether an ensemble of points is randomly distributed or form a clustered distribution. The Ripley function is defined as the mean number of points around a position in a distance less than $r$. When $K(r)$ is larger than the uniform density $\pi r^2$, then points are considered to be clustered. \\
To further sub-segment nanodmain, a Voronoi tessellation can be used: it consists in a partition of the domain based on a set of points $\{p_1, ..., p_n\}$ on a two-dimensional plan. The Voronoi cell $R_k$ is the ensemble of points, whose distance to $p_k$ is smaller or equal to its distance to any other partition point.
By considering the location where the average localization density is for example twice the one obtained by considering the region inside a larger area such as the neuronal contour if our interest is a dendritic spine, it is possible to segment a cluster (Fig.~\ref{figureNanodomains}C-H). This method provides an advantage for defining a cluster with no a priori parameter selection.\\
The algorithm presented in\cite{levet2015sr} allows an automatic segmentation and quantification of protein organization such as AMPARs \cite{levet2015sr} or the nanoscale organization of calcium voltage channel at neuromuscular junction \cite{ghelani2023interactive} (Fig.~\ref{figureNanodomains}A-I). The Voronoi tesselation accounts for more complex shapes of nanodomains compared to ellipses used for potential wells (Fig.~\ref{figureNanodomains}I), but the method cannot be used to estimate the stability of a nanodomain to retain trajectories, as it disregards the dynamics. Future improvements could be to compare tesselations over time, which would require comparing local partitions.
\subsection{Potential wells estimation based on Bayesian inference}
Potential well analysis \cite{hoze2012heterogeneity} has further been confirmed by Bayesian inference of diffusion and force field maps. The biophysical model of the bayesian inferences starts with the overdamped Langevin's equation (eq.~\eqref{LSE12}) and uses maximum a posteriori optimization of the transition probability between pairs of successive points $P(\X(t)|\X(t'))$~\cite{el2016primer} given by
\beq
\frac{dP(\X(t)|\X(t'))}{dt} = -\nabla \cdot \left(-\frac{\nabla U(\X)}{\gamma(\X)}P(\X(t)|\X(t')) - \nabla(D(\X)P(\X(t)|\X(t')))\right),
\eeq
with $t' < t$, and $\gamma(\X)$ is a spatially varying friction coefficient and $U$ the potential energy function. This approach is thus much more computational involved and sub-optimal compared to the classical empirical estimators (discussed and reviewed in \cite{hoze2012heterogeneity,Hoze2017}). While the effective diffusion coefficient could be spatially dependent (to account for crowding), the friction coefficient should in principle be constant in the medium cytoplasm or membrane. Bayesian inferences can be used to estimate parameters on different meshing types such as square grid or Voronoi tesselation. However, they require augmentation data by stochastic simulations to determine the correct initial optimization parameters and do not provide an estimation for the boundary of the well, which is a key element to determine the well energy.
Tools implementing such methods include InferenceMap \cite{beheiry2015inferencemap}, an interactive software package that uses Bayesian method to spatially map the dynamics of SPTs and Tramway~\cite{laurent2022tramway}, a Python library for Bayesian analysis of SPTs.
\subsection{Stability of nanodomains analyzed by time lapse analysis}
\begin{figure}[http!]
	\begin{center}
		\includegraphics[scale=0.8]{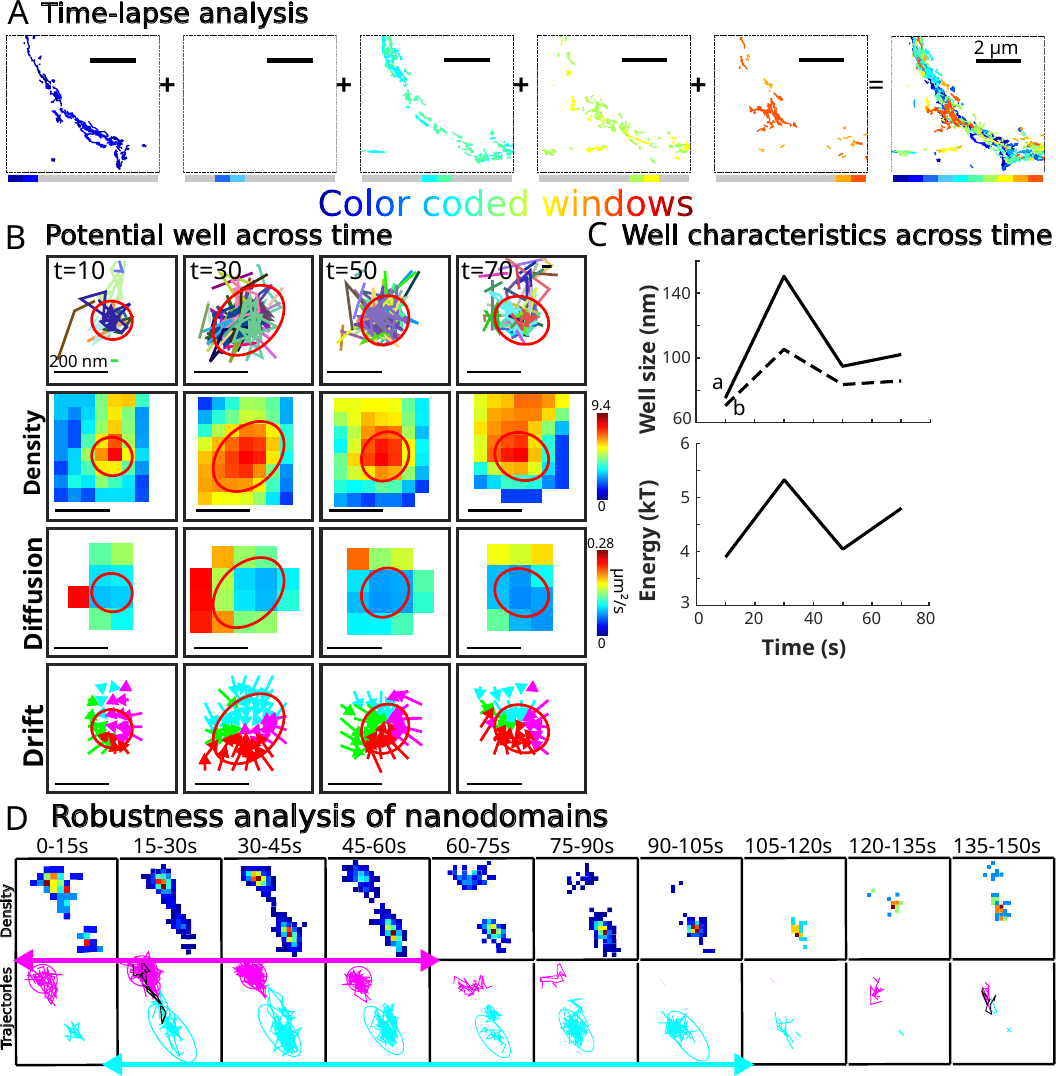}
		\caption{ {\bf Time-lapse analysis to reveal stability of nanodomains.} {\bf (A).} Time-lapse decomposition: trajectories are partitioned into consecutive time windows (color coded) according to the time at which they appear. {\bf (B).} Reconstruction in 20s time windows showing trajectories with elliptic boundary, the density and drift maps. {\bf (C).} Well size and energy over time. {\bf (D).}  Time-lapse analysis showing two neighboring well with different duration colored in purple and cyan respectively.}
		\label{figureTimelapse}
	\end{center}
\end{figure}

Another characterization of nanodomains concern their stability over time: they could be present most of the time or could appear transiently. This effect could have a consequence on their capacity to sequestrate proteins. If they appear and disappear over time, how can SPTs be used to estimate this stability?
when a nanodomain is present, it traps trajectories, while if it start losing energy and disappears, trajectories will not be retained anymore. The strength of the potential is thus a measure of the well stability. Consequently, nanodomain stability can be analyzed by temporally separating SPTs over time by using a sequence of sliding time windows (Fig.~\ref{figureTimelapse}A) of various sizes from few to tens of seconds \cite{holcman2015analysis,Hoze2017,heck2019transient}. It was previously shown that nanodomains are transient for the cases of many receptors such AMPAR \cite{hoze2012heterogeneity} or Calcium voltage channels \cite{heck2019transient,ghelani2023interactive}.\\
Trajectory statistics collected in each time window should be uncorrelated, so that the reconstruction of a well in these time-windows ensures a persistence of the nanodomain. When a well disappears in a given time window, this could be the consequences of absent trajectories or simply because the well loses its attractive strength. The second hypothesis is retained when photoactivation has not saturated as the probability to activate molecules is uniform.\\
After data are collected for each time window, the well reconstruction can be performed by any algorithm described below, extracting their boundary, approximated as an ellipse (Box 1 and 2). The presence of a similar potential well at the same place in two successive windows (Fig.~\ref{figureTimelapse}B) at times $t_k$ and $t_{k+1}$ is based on the criteria that the distance between their centers is less than a given distance and their energy is above a given threshold (at least 1.5kT). The ensemble of consecutive times $(t_q,...,t_r)$,  where a well is first detected at time $t_q$ and disappears at time $t_{r+1}$  is used to define the stability duration $\tau=t_r-t_q$.  This analysis allows to follow the size of the small and large elliptic semi-axis lengths of the wells and the associated energy over time (Fig.~\ref{figureTimelapse}C).\\
To conclude, the average time a well is present over multiple time-frames provides an estimation of its stability (Fig.~\ref{figureTimelapse}D). The overall statistics is obtained by averaging over the residence time of many overlapping trajectories.
\section{Extracting the properties of nanodomains with SPT-based algorithms}
In this final section, we describe several algorithms for automatically detecting nanodomains and extracting their parameters: the centers, the boundaries and the local associated energies. These algorithms mix various technical steps: the boundary is identified using a discontinuity in the density or drift maps. Using a stochastic model for the motion, the various biophysical parameters are identified using optimization procedures such Maximum Likelihood Estimator (MLE) to extract the diffusion coefficient and the amplitude of the potential well \cite{HastieElements2009}, as described in Box 1. We first introduce the two main algorithm followed by variants, including a mutliscale procedure that bypasses the choice of a spatial lengthscale, that depends on the dataset.

All the presented algorithms have in common the initialization phase: a first density map is built over the entire field of view of the experiment and the $x\%$ highest-density local-maxima bins are kept as seeds. Then we use iterative methods, specific to each method to try to fit potential well around each seed. Finally, the different iterations are scored, the optimal one is kept and filtering is done to remove regions not corresponding to potential wells.
\subsection{Drift-based algorithm}
Historically, the first potential well detection algorithm developed for SPT data \cite{HozePNAS} was based only on the local statistics of displacements $ \X(t+\Delta t)-\X(t)$. A square grid is built from the center of the high-density region and displacements from many different trajectories falling in each bin are averaged \cite{holcman2015analysis} to approximate the local drift field $\mb{a}(\X)$. This empirical drift map is compared to the expected drift field based on the stochastic process described by eq. \ref{eq:fbmparab} using a least-square method
\beq\label{LSQEf}
Err_{D(P,r)}(A) &=& \inf_{ \{ \lambda_x,\lambda_y \}} \sum_{X_i \in D(P,r)}\| -\nabla U (X_i) - \mb{b}(X_i)\|^2,
\eeq
where $D(P,r)$ is the disk at the center point $P$ and of boundary radius $r$ and  $\nabla U (X_i) =(-\lambda_x (x-\mu_{x}),\lambda_y (y-\mu_{y}))$ with ${\lambda}_x = \frac{2A}{a^2}, {\lambda}_y = \frac{2A}{b^2}$. In this first version, the boundary was determined manually by searching for the disk $D$ minimizing the least-square error.

An improved version was presented in \cite{parutto2019biophysics} relying on an automatic iterative procedure based on the evaluation of the drift field in concentric rings around the center of the region. The circular boundary is automatically determined as the radius at which the norm of the drift field stops increasing. This improved method also works for elliptic boundaries, by adjusting the previous procedure using the ratio $\frac{{\lambda}_x}{{\lambda}_y}$ to renormalize one axis so that the ellipse is transformed into a disk.

Once the boundary is determined, both methods rely on the least-square estimators for the attraction and diffusion coefficients as well as evaluating the quality of the fitted well, as described in boxes 1 and 2.
\subsection{Density algorithm}
The density-based algorithm reconstruct potential wells boundaries and attraction coefficients based only on the static density of points \cite{parutto2019biophysics}. First it constructs the level set of the local density distribution $\Gamma_{\alpha}$ which is the ensemble of all trajectory points falling in bins with a density greater than $\alpha M^*$:
\beq \label{gammaalf}
\Gamma_{\alpha}= \{ \X_i \, \mbox{such that} \, \rho_e(\x)>\alpha M^* \},
\eeq
where $\rho_e$ is the empirical point density, estimated over the bins of a square grid (eq.~\ref{eq:dens}) and $\alpha \in [0, 1]$ is a density threshold. The expected point density distribution inside a well for a constant diffusion coefficient follows the Boltzmann distribution
\beq
\rho (\X) = B \exp \left((\X -\mu^{(\alpha)})^T C^{(\alpha)} (\X -\mu^{(\alpha)})\right),
\eeq
where $B$ is a constant. Estimating the density of points of a truncated Ornstein-Ulhenbeck process allows to recover the center of the well, the covariance matrix and the boundary. For a circular boundary, we use a succession of concentric rings to compute the point density centered around the estimated center $\hat{\m}$ approximated by the empirical estimators restricted to the points in $\Gamma_{\alpha}$:
\beq \label{mualpha}
\hat{\mu}_{\alpha}^{(u)}= \frac{1}{N_p} \sum_{\{k=1, \X_k \in \Gamma_{\alpha} \}}^{N_p} x_k^{(u)},
\eeq
with $N_p$ the number of points in the ensemble $\Gamma_{\alpha}$ and $u=1..2$. The optimal boundary of the well is located at the inflection point of the distribution corresponding to the transition from a Gaussian density to a uniform one \cite{parutto2019biophysics}.\\

For an elliptic boundary, the previous procedure is adapted using the ratio of the two diagonal terms of the covariance matrix $C^{(\alpha)}$ defined as
\beq \label{cov}
\hat{C}_{ij}^{(\alpha)}= \ds{ \frac{1}{N_p-1} \sum_{ \{k=1, \X_k \in \Gamma_{\alpha} \}}^{N_p} X_{i,k}X_{j,k}},
\eeq

The ratio $\sqrt{\frac{\hat{C}_{11}^{(\alpha)}}{\hat{C}_{22}^{(\alpha)}}}$ is used to determine the rescaled distance for which an ellipse can be mapped to a disk (Box 2).
To conclude, the density algorithm can be used to recover the nanodomain boundary. Then the biophysical parameter can be computed using the MLE algorithm (Box 1).
\subsection{Hybrid density-drift}
Using both the distribution of points and their displacements leads to a hybrid density-drift algorithm to estimate the potential well boundary characteristics. Here, a sequence of ellipses are fitted by iteratively growing square regions centered around the local density maximum. For each iteration, a Principal Component Analysis (PCA) is used to extract an $x\%$ confidence interval ellipse from the points falling in that region (Box2). The attraction coefficient $A$ is computed from the score $S$ (eq. \ref{estimA}) and the diffusion coefficient is approximated using the second moment statistics (Box2).\\
It is often difficult to identify the appropriate spatial scale in SPTs data. This is an inherent difficulty as the size of nanodomains is a priori unknown and can vary from tens to hundreds of nanometers, depending on its location, the underlying molecular assembly, the type of organelles and membranes.  The multiscale version of the hybrid algorithms allows to find the well without human intervention. It identifies the optimal spatial scale by detecting potential wells from multiple spatial scales $\Delta x$ (eq.\ref{eq:wellMLED}) \cite{parutto2022high}. This multiscale step improves the accuracy of boundary extraction as well as the other parameters.
\subsection{How to chose the best algorithm when analysing SPTs}
The hybrid-drift algorithm is well suited for large wells ($> 200$ nm) with a sufficient number of trajectories ($> 50$) as its accuracy depends on the number of bins falling inside the boundary of the well. The bins themselves need to be large enough (usually around $20$ to $50$ nm) to contain enough displacements for computing reliable drift values ($5$ to $7$). This binning is a key step to evaluate wells with a size comparable to the bin size as it reduces the accuracy of the parameter estimation, but the extra average layer leads to more stable results for large wells. Note however that with this method, the non-uniform distributions of points inside the well will usually lead to a slight underestimation of the dynamical parameters of the well.\\
The hybrid-MLE algorithm has been developed as an adaptation of the hybrid-drift algorithm for small wells ($< 100 nm$) with low number of trajectories ($< 50$). Contrary to the drift estimation, the MLE estimator allows recovering the dynamical parameters from the ensemble of displacements instead of relying only on local averages. In addition, the boundary evaluation based on computing the maximum likelihood from the displacements added at each iteration improves the sensitivity compared to the other methods based on the accumulation of displacements.\\
Finally the density algorithm recovers well characteristics based mostly on the shape of the local density peak and thus is well suited when trajectories are unreliable. Finding the boundary of the well from the density distribution however uses the inflection point in the distribution which can be quite imprecise with a low amount of data.\\
A general observation for all these methods is that the accuracy of the well boundary is improved when there are sufficient amount of trajectories both inside and outside the well. Indeed the difference in characteristics of both density and dynamics of trajectories between inside and outside the well allows to more precisely determine the boundary. We summarized in table \ref{table:algotouse} the advantages of each algorithm.
\begin{table}[http!]
\begin{center}
\begin{tabular}{p{2.5cm}|p{3cm}|p{3cm}|p{3cm}}	
      & Density & Drift & Hybrid\\\hline
Advantage &  {No trajectories needed, use for unreliable trajectories} & Most stable method with sufficient data. Use for large wells& Do not need a large amount of data, use for small wells \\\hline
Inconvenience &   Require a large number of points  & Require optimal conditions& Can be unstable\\\hline
\end{tabular}
\end{center}
\label{table:algotouse}
\caption{\textbf{Advantages and inconvenient of the different well detection methods.}}
\end{table}
The present methods and algorithms have been implemented into an imageJ plugin called "TrajectoryAnalysis". The plugin allows reconstructing various maps (trajectories, density, drift, diffusion), detect potential wells and reconstruct the graph associated to trajectories as well as automatically reconstructing the potential wells contained in a dataset.
\section{Concluding remarks and perspectives}
The past decade has seen significant developments in both the experimental acquisition of high-throughput single-particles trajectories as well as their analysis based on statistical methods, stochastic modeling and the construction of optimal estimators and algorithms. As reviewed here, these approaches allow to identify high-density regions and to compute key biophysical parameters such as the diffusion coefficient, the energy of the wells and their specific boundaries. There are multiple algorithms based on Bayesian approaches \cite{masson2014} and other based on optimal estimators \cite{hoze2012heterogeneity,parutto2022high}. However, automated detection is another improvement to account for large data sets. Indeed, multiple improvements arise throughout the past years: while density or diffusion map resolution was of the order of 100-200nm, these new algorithms allowed to gain an order of magnitude using a sliding disk analysis \cite{parutto2022high}. In the future, it is could be possible to identify other drift patterns, apart from local convergence. Drift map could reveal the emergence of directional paths.
Another extension of the reconstruction algorithm could be to compute an anomalous diffusion map where the anomalous exponent is computed at each point in order to locally characterize non-diffusive motion \cite{sokolov2012}.\\
Further, nanodomains could be analysed while measuring the local membrane characteristics: membrane-membrane contacts, elasticity and bending would be key to relate nanodomain stability with membrane organization. Possibly nanodomain stability could be correlated to long-lasting membrane contacts. Nanodomain formation could be a coarse-grained signal serving for membrane interaction between compartments such as plasma and organelle membranes. Compartment organization such ER-plasma membrane or mitochondrial membrane is common to all cell types, including neurons with a strong polarization of axons and dendrites, but is also relevant for glial cells showing a large diversity of ER-PM contacts, a key regulator of store-operated calcium entry in the ORAI-STIM connection \cite{basnayake2021}.\\
Another extension of the present algorithms would be to account for three-dimensional SPTs and to reconstruct the explored membrane, by combining trajectories to increase the coarse z-resolution and also recovering the biophysical properties. These experimental and theoretical challenges would depend on ultra-lightsheet microscope. Future dual or multiplexed color imaging could also allow to study statistical interactions between molecular partners, their interacting forces and the co-localization of their high-density nanodomains. It will also be interesting to extract transient compartments with different time scales, a key advantage of SPT versus point density.\\
Nowadays, both the biophysical theory and softwares implementing them are available to precisely analyze SPT data, helping disseminate these techniques. They could be generalized to handle 3-dimensional data, multiple populations or work with non-convex domains by including a tesselation. Integrating three dimensional membrane reconstruction would allow to correlate nanodomain dynamics with membrane curvature and thus better clarify how force tension signaling could be transformed into nanodomain molecular assembly signaling.

\begin{figure}[http!]
\begin{center}
\includegraphics[scale=0.4]{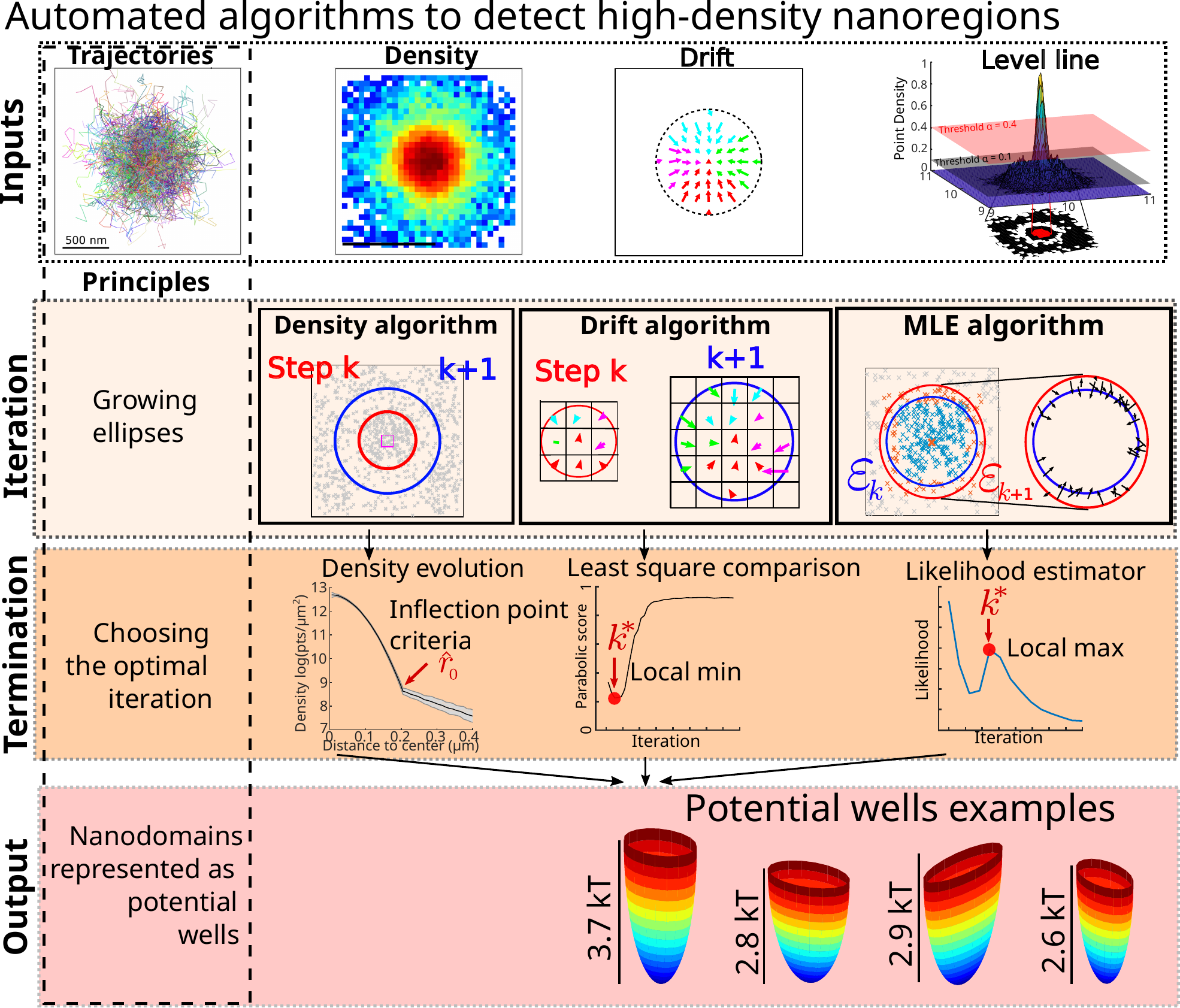}
\caption{ {\bf Box2: Specific steps of the main algorithms (density, Drift and Hybrid) to reconstruct high-density nanodomains as potential wells.} {\bf (A).} Inputs are trajectories, density, drift and level line.  {\bf (B).} Iteration step consisting in increasing the considered neighborhood around the high-density center {\bf (C).} Termination and estimation procedure: the optimal ellipse is estimated as the inflection point for the density algorithm, the minimization of the LSQE or as the local maximum for the hybrid algorithm using the MLE estimated on rings {\bf (D).} Outputs consist in the reconstructed well geometry, diffusion $D$ and attraction $A$ coefficients.}
\label{figureWellsMethods}
\end{center}
\end{figure}
\newpage
\section{BOX1: Optimal estimation of biophysical parameters for nanodomains}
In this box, we describe statistical estimators used to recover the potential well parameters. A parabolic well is characterized by seven parameters falling into two categories:\\
{\noindent \bf Geometrical}
\begin{enumerate}
  \item Center $(\mu_x, \mu_y)$
  \item Large and small semi-axis $a,b$ lengths of the elliptic boundary $\mathcal E$ and the ellipse orientation $\phi$.
\end{enumerate}
{\bf \noindent Dynamical:}
\begin{enumerate}
  \item Diffusion coefficient $D$ inside the well
  \item Attraction coefficient $A$ (see eq. \ref{eqA}).
\end{enumerate}
We now present three different methods for estimating for the attraction and diffusion coefficients.\\
{\noindent \bf Least Square Quadratic Estimator (LSQE).} This method relies on the characteristics of the well energy function
\beq\label{parabU}
U(\X) = \left\{ \begin{array}{ll}
	A\left[\left(\frac{(x - \mu_x)}{a}\right)^2 + \left(\frac{(y - \mu_y)}{b}\right)^2\right] & \hbox{if } \X \in {\mathcal E}\\
	1                                                               & \hbox{otherwise}
\end{array} \right. ,
\eeq
where ${\mathcal E}=\{ (x,y) \hbox{ such that } A\left[\left(\frac{(x - \mu_x)}{a}\right)^2 + \left(\frac{(y - \mu_y)}{b}\right)^2\right] \leq 1\} $ is the basin of attraction.  The resulting drift field inside the well is given by
\beq \label{well:driftfield}
\nabla U(\X) = -2A
\left[\begin{array}{ll}
	\ds \frac{x - \mu_x}{a^2} \\
	\ds \frac{y - \mu_y}{b^2}
\end{array} \right].
\eeq
The parameter A can be obtained from trajectory data by minimizing the square error between the analytical expression eq.~\ref{well:driftfield} and the drift field $\vec{\f}(\X)$ \cite{HozePNAS}
\beq \label{mini}
Err(A^{*}) &=& \min_{a,b} \int_{ \X \in \mathcal E} \|\nabla U (\X) +\vec{\f}(\X)\|^2 {dS_{\X}}.
\eeq
The discretized version uses a drift map constructed over a square grid centered at $\X_{1},..,\X_{N}$ is
\beq\label{eq:leqERR}
Err_{N}(A^*) &=& \frac{1}{N} \sum_{i=1}^{N}\| -\nabla U (\X_i) - \vec{\f}(\X_i)\|^2 =\sum_{i=1}^{N} \left(f_x(\X_i) +2A\frac{x_i}{a^2}\right)^2+\left(f_y(\X_i) +2A\frac{y_i}{b^2}\right)^2,\nonumber\\
\eeq
where  $\vec{f}(\X_i) = (f_x(\X_i),f_y(\X_i))$ is the discretised vector field on the square centered at points $\X_i$. The value of $A^*$ that minimises \ref{mini} is given by
\beq \label{estimA}
A^* = -{\ds \frac{1}{2} \frac{\sum_{i=1}^{N} \frac{f_x(\x_i) x_i}{a^2} + \frac{f_y(\x_i) y_i}{b^2}}{\sum_{i=1}^{N} \frac{x_i^2}{a^4} + \frac{y_i^2}{b^4}}}.
\eeq
Similarly, the diffusion coefficient $D$ is computed from the ensemble of $N$ trajectories of size $N_i$ points, $\X_i(t_j)=[x_i^{(1)}(t_j), x_i^{(2)}(t_j)]$, $i=1..N$, $j=1..N_i$ using the following estimators for all trajectory displacements starting inside the ellipse $\mathcal E$
\beq
D_{uv}(\mathcal E) &\approx& \frac{1}{N_k}\sum_{i=1}^{N}\sum_{j=0, \X_i(t_j)\in \mathcal E}^{N_i-1}\frac{(x_i^{(u)}(t_{j+1})-x_i^{(v)}(t_j))^2}{2\Delta t},
\eeq
where $N_k$ is the number of displacements falling in  $\mathcal E$. The estimated value $\tilde D$ is the average of the diagonal term of the tensor
\beq \label{diffcoef}
\tilde D(\mathcal E) = \frac{D_{11}(\mathcal E) + D_{22}(\mathcal E) }{2}.
\eeq
{\bf \noindent Maximum Likelihood Estimator (MLE)}
For dynamics inside a potential well, the MLE relies on a truncated Ornstein-Uhlenbeck process
\beq \label{1dequation}
\dot{\X} =
\left\{\begin{array}{ll}
	-\llll(\X(t) - \m)+ \sqrt{2D} \dot W & X \in \mathcal E\\[7pt]
	\sqrt{2D} \dot W  & \textrm{ otherwise}.
\end{array}
\right.,
\eeq
where $\llll = [\lambda_x, \lambda_y] = [\frac{2A}{a^2}, \frac{2A}{b^2}]$ and $W$ is a white noise. The probability density function of observing $\X(t')$ given  $\X(t)$ with $t - t' = \Delta t > 0$ is given by 2D-Gaussian with mean
\beq
\mm(\X(t')) &=& \X(t)e^{-\llll \Delta t} + \m(1 - e^{-\llll \Delta t})
\eeq
and standard deviation
\beq \label{eq:s}
\s(t) &=& \frac{\sigma^2(1 - e^{-2\llll \Delta t})}{2\llll},
\eeq
where $\sigma = \sqrt{2D}$. The log-likelihood function of observing the successive displacements $(\X_i(t_j),\X_i(t_{j+1}))$, $i=1 \ldots N$, from observed trajectories is given by
\beq\label{eq:welllikelihood}
\mathcal{L}(\m, \llll, \s|\X_1,..,\X_n) &=& \sum\limits_{i=1}^N \log(p(\X_i(t_{j+1}), \X_i(t_j))\nonumber\\
&=& -\frac{1}{2} \sum\limits_{i=1}^N \left[\log(2\pi \s) + \frac{(\X_i(t_{j+1}) - \mm(\X_i(t_j)))^2}{\s} \right],
\eeq
from which we get the two estimators ~\cite{tang2009} (zero derivative with respect to $\lambda$ and $\mu$ and $D$)
\beq\label{eq:wellMLElambda}
\tilde\llll_{N} &=& -\frac{1}{\Delta t} \log \left( \frac{\left(\sum\limits_{i=1}^N \X_i(t_{j+1})X_i(t_j)\right) - \left(\frac{1}{N} \sum\limits_{i=1}^N \X_i(t_j)\right)\left(\sum\limits_{i=1}^N \X_i(t_{j+1}) \right)}{\left(\sum\limits_{i=1}^N \X_i(t_j)^2 \right) - \frac{1}{N}\left(\frac{1}{N} \sum\limits_{i=1}^N \X_i(t_j))\right)^2} \right),\nonumber\\
\eeq
and
\beq\label{eq:wellMLED}
\tilde D_{N} &=& \frac{\llll}{N(1 - e^{-2\llll \Delta t})} \sum\limits_{i=1}^N [\X_i(t_{j+1}) - \mm(\X_i(t_j))]^2.
\eeq
{\bf \noindent Density Estimator}
The steady-state density distribution of an Orstein-Uhlenbeck process is given by
\beq
\rho(\X) = N_0 \exp\left\{-\frac{A}{D}\left[\left(\frac{x-\mu_x}{a}\right)^2 + \left(\frac{y-\mu_y}{b}\right)^2\right]\right\},
\eeq
wich corresponds to a Gaussian distribution of center $[\mu_x, \mu_y]$ and covariance matrix
\beq
C = \frac{A}{D}\left[\begin{array}{cc}
	\frac{1}{a^2} & 0\\
	0 & \frac{1}{b^2}
\end{array}
\right].
\eeq
Inverting this equation leads to an estimate for $A$ based on the diffusion coefficient and the covariance matrix
\beq
\tilde{A}_{1,1} = \frac{Da^2}{C_{1,1}}\nonumber\\
\tilde{A}_{2,2} = \frac{Db^2}{C_{2,2}}.
\eeq

\newpage
\section{BOX2: Algorithms to reconstruct the boundary of nanodomains}
The estimators presented in Box 1 rely on the correct determination of the well boundary $\mathcal E$. This box 2 presents different algorithms to estimate the boundary. These algorithms consists of three steps: initialization to find high-density regions, iterations to increase successive ellipses and a final step to evaluate the optimal boundary is determined. We start by fixing the maximum domain in  which the boundary can be located, equivalently, we fix a maximum iteration number $k_{max}$\\

{\bf \noindent Initiation.} To find high density regions in the field of view, a grid $G_{\Delta x}$ is used with square bins of size $\Delta x$ from which we compute the density map
\beq\label{eq:dens}
\rho_{\Delta x}(\X_k) = \frac{N_k}{(\Delta x)^2},
\eeq
where $N_k$ is the number of trajectory points falling into the bin centered at $x_k$. The highest $d\%$ (a parameters of the method usually 5\%) bins from $\rho_{\Delta x}$ is selected as possible regions containing a potential well. The center is the most dense bin $\m$.\\

{\noindent \bf Iterations.} For each selected high-density region, the iteration steps consist in computing an ellipse to approximate the boundary centered at the maximum of the density. This iteration can be made with
several algorithms:
\begin{itemize}
	\item {\bf Density algorithm.} The density method considers increasingly concentric annuli centered at $\m$. When considering a circular boundary, we compute the density of point $N(r_k)$ by counting the number of point falling in the annulus of radius $r_k$, with $k=1..k_{max}$. For an elliptic boundary, we first determine the ratio $\kappa$ that maximizes $C_v(r_k)=\sqrt{a_k / b_k}$ for $k=1..k_{max}$. This parameter is used to define a distance function
	\beq
	r_e(\X) = \sqrt{\left(x^{(1)} - \mu_0^{(1)}\right)^2 + \kappa \left(x^{(2)} - \mu_0^{(2)}\right)^2},
	\eeq
	that locally transforms the elliptical density distribution into a circular one with the same center. Finally, we compute the local density distribution $N_{e}(r_k)$ by counting the number of points falling in the annulus of radius $r_k$ based on the modified distance $r_e$.
    \item {\bf Hybrid:} The hybrid method considers increasingly larger square neighborhoods around the initial center. For each iteration $k=1..K$, we keep only the points contained inside the square $\Gamma_{k,\Delta x}$ of size $[(2k+1)\times (2k+1)] (\Delta x)^2$ and centered at the center of mass. Based on these points, we compute the $p\%$ confidence ellipse based on principal component analysis of the covariance matrix from which we obtain the elliptical semi-axes $a_k$, $b_k$ matrix $C_k$ the ellipse orientation angle $\varphi_k$.
\end{itemize}

{\noindent \bf Termination and scoring.} The optimal boundary is estimated as the ellipse that optimizes the score associated to a specific algorithm:
\begin{itemize}
	\item {\bf Density:} we select the minimal iteration $k^*$ such that the derivative with respect to the radius has a discontinuity: $|N_e(r_k^*)|- |N_e(r_{k^*-1})|>T$, where $T$ is a threshold to fix.   It is the first iteration where the derivative of the density with respect to the distance to the center stops decreasing. The selected ellipse is $S_{k^*}$.
	\item {\bf Hybrid-Drift:} The error is computed by minimizing the parabolic index $S$ derived from the least-square error formula presented in eq.~\ref{eq:leqERR} in Box 1,
	\beq \label{scoreexpression}
	S_k(a_k) = 1 - \frac{\left( \sum\limits_{i=1}^M \frac{f_k^{(1)}(\x_i)x_i^{(1)}}{a^2} + \frac{f_k^{(2)}(\x_i)x_i^{(2)}}{b^2} \right)^2}{\left( \sum\limits_{i=1}^M \frac{(x_i^{(1)})^2}{a^4} + \frac{(x_i^{(2)})^2}{b^4} \right) \left(\sum\limits_{i=1}^M ||\mb{f}_k(\x_i)||^2 \right)},
	\eeq
	where there are $M$ square bins centered at $\x_i = (x_i^{(1)}, x_i^{(2)})$ falling inside the well boundary, $\vec{f}$ is the drift field and $a,b$ are the ellipse semi-axes. The index $S_k \in [0, 1]$ is such that $S_k \rightarrow 0$ for a drift field generated by a parabolic potential well and $S_k \rightarrow 1$ for a random drift vector field.

	We select the iteration $k^*$ that minimizes  $S$: $k^* = \arg\min_{k=1\ldots K} S_k(a_k, A_k)$. We estimate the diffusion coefficient inside the well using the local estimator (eq.~\ref{diffcoef}) and the coefficient from formula \ref{estimA} for all the displacements inside the ellipse $\varepsilon_{k^*}$.
	
	\item {\bf Hybrid-MLE:} Two variants of this algorithm exists:
	\begin{itemize}
		\item {\bf Cumulative MLE:} for each iteration $k$, we compute the likelihood (eq. \ref{eq:welllikelihood}, BOX 1) associated to all the displacements falling inside the ellipse $\varepsilon_{k}$. The best iteration $k^*$ is the one that globally maximizes the likehood.
		\item  {\bf Boundary MLE:} for each iteration $k$, we compute the likelihood (eq. \ref{eq:welllikelihood}, BOX 1) associated only to the displacements falling in-between the ellipses  $\varepsilon_{k-1}$ and  $\varepsilon_{k}$. The best iteration $k^*$ in this case is obtained as the second local maximum~\cite{parutto2022high}. This procedure is more sensitive when small number of trajectories are available.
	\end{itemize}
	The diffusion coefficient $D$ and parameter $A$ are computed from the MLE procedure applied for all the displacements falling inside the ellipse $\varepsilon_{k^*}$.
\end{itemize}
We summarize the algorithm as follows:
\begin{algorithm}
\caption{Well Detection Algorithm}
\KwData{$trajs$ an ensemble of trajectories.}
\KwData{$\Delta x$ bin size ($\mu m$) for initial density map.}
\KwData{$x\%$ percentage of high-density regions to keep.}
\KwData{$params$ method-specific parameters.}
\KwResult{$wells$ an ensemble of wells.}
\Begin
{
	$grid \gets \textrm{generate\_grid\_over\_fov}(\Delta x)$\;
	$dens \gets \textrm{compute\_density\_map}(trajs, grid)$\;
	$seeds \gets \textrm{select\_high\_bins}(dens, grid, x\%)$\;
	\For{$seed \in seeds$}
	{
		$ellipses \gets \textrm{grow\_ellipses}(seed, params)$\;
		$\varepsilon^* \gets \textrm{select\_optimal\_ellipse}(ellipses)$\;
		$A^* \gets \textrm{compute\_A\_coefficient}(trajs, \varepsilon^*)$\;
		$D^* \gets \textrm{compute\_D\_coefficient}(trajs, \varepsilon^*)$\;
		$w \gets (\varepsilon^*, A^*, D^*)$\;
		\If{$\textrm{validate\_well}(w)$}
		{
			$wells \gets wells \cup w$\;
		}
	}
}
\end{algorithm}

\newpage
\normalem
\bibliographystyle{ieeetr}
\bibliography{references10final-dh,ref_general,biblioCRM22,MaterialAndMethodsBibliography3,RMPbiblio4newN2,biblioPierre}
\end{document}